\documentclass{aastex}
\usepackage{emulateapj5}
\usepackage{onecolfloat}

\newcommand{\beq}{\begin{equation}}
\newcommand{\eeq}{\end{equation}}

\newcommand{\be}{\begin{equation}}
\newcommand{\ee}{\end{equation}}

\newcommand{\hmpc}{h^{-1}\mathrm{Mpc}}

\newcommand{\hMsun}{h^{-1}M_{\odot}}

\newcommand{\band}[2]{\ensuremath{^{#1}\!{#2}}}
\newcommand{\Mr}{M_{\band{0.1}{r}}}

\newcommand{\dg}{$^\circ$}
\newcommand{\hinv}{$h^{-1}$}
 
\bibpunct[,]{(}{)}{;}{a}{}{,}
\begin{document}
 
\twocolumn[
\title{Foreground and Source of a Cluster of Ultra-high Energy Cosmic Rays}
\author{
Glennys R. Farrar, \altaffilmark{1}
Andreas A. Berlind, \altaffilmark{1}
David W. Hogg, \altaffilmark{1}
}
\affil{Center for Cosmology and Particle Physics,
Department of Physics\\
New York University,
New York, NY 10003, USA}

\begin{abstract}
We investigate the origin of a nearly pointlike cluster of 5 ultrahigh energy cosmic rays at RA $\approx 169.2^\circ$ and dec $\approx 56.8^\circ$, using Sloan Digital Sky Survey and other data.   No particular source candidates are found near the estimated source direction, but the direction is exceptional in having a likely merging pair of galaxy clusters at $140\hmpc$, with an unusually low foreground density.  Large scale shocks or another product of the merging galaxy clusters may accelerate the UHECRs, or the merging galaxy clusters may be coincidental and the UHECRs may be accelerated in a rare event of an unexceptional progenitor.  Low magnetic deflections in the foreground void may explain why this is the only identified pointlike cluster of so many UHECRs.   
\end{abstract}

\keywords{cosmic rays, cosmology: large-scale structure of universe}
]
                                                                                
\altaffiltext{1}{gf25@nyu.edu, aberlind@cosmo.nyu.edu,
david.hogg@nyu.edu}

\maketitle

\section{Introduction} \label{intro}

Recently,  a cluster of 5 ultra-high energy cosmic ray (UHECR) events was found in the complete published AGASA-HiRes dataset, whose distribution of arrival directions is consistent with their having a common pointlike origin and little magnetic smearing \citep{gfclus}.  The cluster of four highest energy events was identified in \citet{HRGF}, in the dataset consisting of 57 AGASA events above 40 EeV and 37 HiRes events above 30 EeV.  The fifth event joins the others when the full HiRes dataset above 10 EeV is used, adding 214 events to the 94-event high energy dataset containing the quad.  The ({\it a posteriori}) probability that the quadruplet is a chance occurrence in a random isotropic dataset of 94 events is $ 2 \times 10^{-3}$ \citep{gfclus}, motivating a search for a common source.  In this work we address the question of what objects offer greatest promise for being or containing the source, and whether the large-scale structure of the environment in the direction of the cluster gives any clue about the likely distance of the source and/or the reason the magnetic dispersion among the 5 events is so low.   

The highest energy event in the cluster was observed by AGASA, who measured its energy to be 77.6 EeV \citep{AGASAupdate}, with an estimated $\pm 25\%$ statistical and $\pm 18\%$ systematic energy uncertainty \citep{AGASAenergy}.  Due to energy loss during propagation, especially from photopion scattering on the CMB, the maximum distance for the source of this UHECR cluster can be inferred to be $\sim 210$ Mpc at 99\% CL using the nominal energy of the highest energy event, 77.6 EeV, or $\sim 430$ Mpc for a 30\% lower energy of 54.3 EeV (S. Balberg, private communication in \citet{gfclus}).  The fifth event gives an essentially perfect fit to the cluster hypothesis with gaussian magnetic dispersion $\sim E^{-1}$, but given the large number of added events when the HiRes threshold is reduced from 30 EeV to 10 EeV, there is a 1 in 6 chance that the 5th event is a coincidental association.  The source direction was reconstructed in \citet{gfclus}, using just the four highest energy events and using all five events, and with and without magnetic dispersion and deflection.  For definiteness we adopt the source direction obtained with all 5 events and magnetic dispersion: \{169.2\dg, 56.8\dg \} with a $1.2^\circ$ 99\% error radius; other combinations barely change the RA but some reduce the dec by up to $ 0.4^\circ$ \citep{gfclus} which is well-within the uncertainty.

There are two natural explanations for why this cluster of 4-5 UHECRs is unique in the northern hemisphere in having a spatial dispersion consistent with being pointlike, for such a large number of events:
\begin{itemize} \item 
This source is the closest and/or most powerful one, and other sources have a lower flux at Earth. 
\item 
Other sources may have as large or larger flux, but larger magnetic dispersion in their foregrounds prevents their clusters from being reconstructed unambiguously.
\end{itemize} 
In order to help choose between these options, we examine the observations to see if there is any inherent evidence that the magnetic field may be lower here than in other directions.   Serendipitously, the cluster lies in the region covered by SDSS Data Release 3 \citep{sdssDR3}, which we use to characterize the environment out to $300 h^{-1}$ Mpc.  Besides this cluster, AGASA and HiRes observe six doublets\footnote{In conference proceedings, speakers from AGASA also refer to a second triplet, in addition to the one that was promoted to a quadruplet by HiRes, but its coordinates are not available.}:  5 with AGASA alone and one which is an AGASA-HiRes pair \citep{HRclus04}.  Unfortunately, only the quad falls in the part of the sky covered by
the SDSS DR3.  We have tried using other surveys, such as the all-sky Point Source redshift Catalog (PSCz, \citealt{PSCz}), but they are not sensitive enough to resolve the structure adequately for our purpose, so we cannot perform the analogous test for those doublets.   

We investigate the large-scale structure of matter in the direction of the UHECR 
quad using two volume-limited samples of galaxies constructed from the SDSS DR3: 
one that is complete down to an $r$-band absolute magnitude of $\Mr<-20$ and goes
from $45\hmpc$ out to $300\hmpc$, and one that is complete down to $\Mr<-18.3$ and 
goes from $45\hmpc$ out to $150\hmpc$.  Absolute magnitudes are k-corrected to a 
redshift of 0.1 and are provided by the New York University Value-Added Galaxy Catalog
\citep{NYUVAGC}.  In addition to studying the overall density field, we also look for 
galaxy clusters, which are the largest gravitationally bound objects in the universe.
We identify galaxy clusters in our larger-volume volume-limited sample using a simple 
friends-of-friends algorithm with perpendicular and line-of-sight linking lengths of 
0.14 and 0.7 times the mean inter-galaxy separation, respectively, and we only retain 
clusters containing 10 or more member galaxies and a minimum estimated mass of 
$10^{14}\hMsun$.  We make rough mass estimates for the clusters by assuming a monotonic 
relation between cluster mass and luminosity, and matching the cluster luminosity 
function to the theoretical mass function of dark matter halos for a concordance
cosmological model.

\section{Results and Discussion} \label{results}

\begin{figure}[h]
\epsscale{1.0}
\plotone{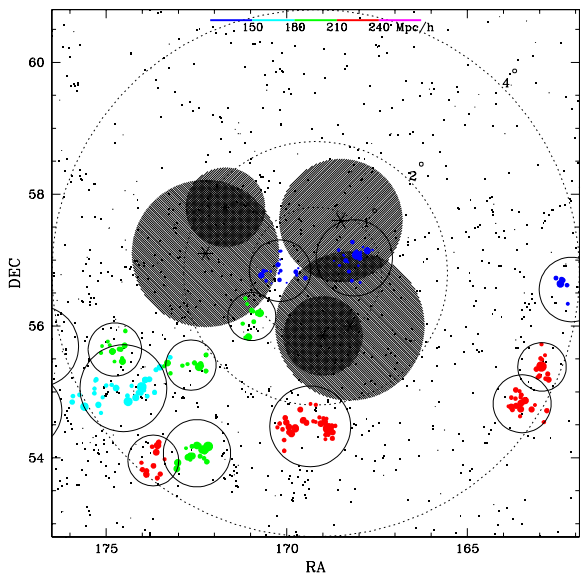}
\caption{
$8^{\circ}\times 8^{\circ}$ field centered on the best-fit UHECR cluster direction 
\{169.2\dg, 56.8\dg \}.  Starred symbols show the arrival directions of the five UHECR 
events and shaded circles show their ``$1\sigma$" error domains, i.e., the regions 
expected to contain the true arrival direction in 68\% of comparable measurements.
The size of the stars is proportional to the energy of the UHECR events.  Dots represent
the positions of galaxies in a volume-limited sample constructed from the SDSS redshift 
survey (sample described in text).  Galaxies that are members of clusters are shown as
dots with sizes proportional to their luminosities and color-coded according to their 
redshift.  Circles indicate the estimated cluster virial radii.
}
\label{f1}
\end{figure}

\begin{figure}[h]
\epsscale{1.0}
\plotone{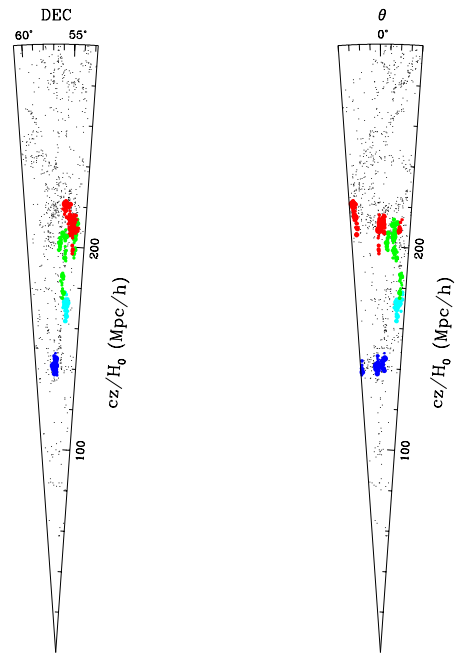}
\caption{
Side-views of the galaxies shown in Fig~1.
} 
\label{f2}
\end{figure}

Figure~\ref{f1} shows SDSS galaxies from our $\Mr<-20$ volume-limited sample in an $8^{\circ}\times 8^{\circ}$ field centered on \{169.2\dg, 56.8\dg \}.  The starred symbols and grey disks show the individual UHECR events with their ``$1 \sigma$" domains.  Galaxies in clusters are shown as dots with sizes proportional to their luminosities, colored-coded to show their distances, with the estimated virial radii of the galaxy clusters indicated by circles. The same field is shown in orthogonal slices in Figure~\ref{f2}. The most striking feature in these figures is the pair of galaxy clusters at $140\hmpc$.  Their physical proximity to each other suggests that they are either currently merging or will likely merge within a dynamical time.  We return to this merging pair later.

\begin{figure}[h]
\epsscale{1.0}
\plotone{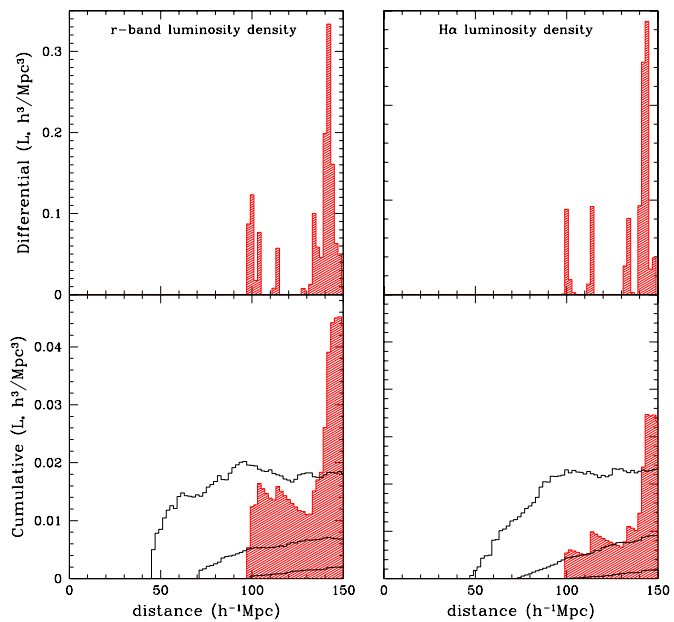}
\caption{The upper-left panel shows the $r$-band luminosity density (measured 
using the smaller-volume of our two volume-limited samples) as a function of 
distance in the direction of the UHECR cluster, averaged over a disk of 1\dg 
radius, out to $150\hmpc$.  The shaded histogram in the bottom-left panel shows 
the corresponding cumulative luminosity density; for comparison, the central 
unshaded histogram shows the median cumulative luminosity density in the SDSS 
DR3, computed from 688 independent lines-of-sight, while the upper (lower) 
histograms show the values such that only 10\% of the cases have higher (lower)
integrated luminosity.  The luminosity density in all lines-of-sight drops to 
zero below $45\hmpc$ because that is the lower limit of our volume-limited sample.  
The right-hand-side panels are the same, using H$\alpha$ rather than $r$-band 
luminosity.
} 
\label{f3}
\end{figure}

The magnetic field strength can be expected to be highest in regions of high temperature and pressure, and lowest in voids.  The luminosity density is thus a convenient albeit approximate surrogate for local magnetic field strength, and we adopt it as a tool to compare the environment in the direction of the cluster to that of average SDSS fields.  Figure~\ref{f3} reveals that in the direction of the UHECR cluster there is
very little matter out to $100\hmpc$.  At $100\hmpc$, only 11.5\% of 
directions in the full SDSS DR3 have as low or lower cumulative $r$-band 
luminosity density as the UHECR direction.  Evidently, if the source is in 
the density enhancement at $100\hmpc$, it is plausible that the magnetic 
deflections experienced by UHECRs en route to Earth may be much smaller
than for sources in other, more typical directions.
The same conclusions hold when we compute the luminosity density in
cones of 2\dg radius, rather than 1\dg (not shown).  This means that
the properties we deduce about the UHECR line-of-sight are robust over the
region of uncertainty of source location and magnetic dispersion. 

We now turn to the question of what the source or accelerator of the UHECRs might be, and what can be deduced about its likely distance.  Some candidate UHECR sources (e.g., GRBs) are found preferentially in star-forming regions.  We can probe directly the rate of star formation in our field by repeating the analysis described above for $r$-band luminosity using H$\alpha$ luminosity.  We use the H$\alpha$ luminosities computed by \citet{Halpha}.  The results are given in the right-hand-side panels of Figure~\ref{f3}. They show that the high matter-density regions along the UHECR line-of-sight have enhanced star-formation, roughly in proportion to the matter density.  

Figure~\ref{f3} shows that, beginning abruptly at 100\hinv Mpc, this field has above average luminosity column-depth.  Indeed, just beyond the merging clusters, the cumulative $r$-band (H$\alpha$) luminosity for the 1\dg field is greater than in all but $\sim 0.6\%$ (2.9\%) of the trials, gradually decreasing at larger distances but remaining in the top $\sim 5\%$ out to 300 \hinv Mpc.  

Merging galaxy clusters offer multiple acceleration options. One possibility, not previously explicitly considered to our knowledge, is that the UHECRs could be accelerated in large scale shocks produced by the merging galaxy clusters.  Field strengths several times as large as those produced by infall onto an isolated cluster of comparable mass would be expected.  Evidence for such shocks has been detected in the Chandra observation of the high-redshift galaxy cluster Cl J0152.7 1357 \citep{chandraMergClus}.  Furthermore, when galaxy clusters merge, the merger rate of their constituent galaxies would be expected to increase.  This in turn should stimulate star-formation and therefore increase the rate of GRBs and magnetar births, both proposed as UHECR accelerators (see, e. g., \citealt{waxmanUHECR,aronsUHECR}).  We looked at the morphology of individual galaxies in the merging clusters to estimate the rate of galaxy mergers, but find that an enhancement in comparison to field galaxies, if present, is not statistically significant.   Galaxy merging could also activate quiescent AGNs by perturbing their accretion disks.   By contrast, undisturbed galaxy clusters presumably have a lower star formation rate per unit mass because of the deficit of cold gas, although gravitational infall on rich clusters should produce large scale shocks.  
\begin{figure}[h]
\epsscale{1.0}
\plotone{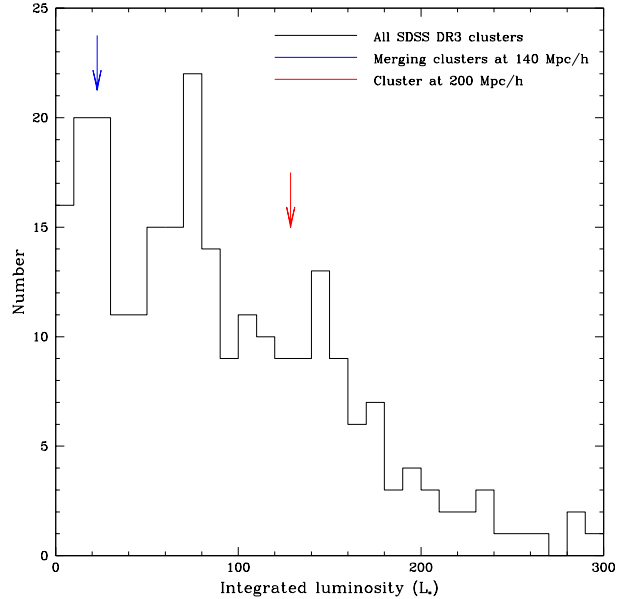}
\caption{
Distribution of the cumulative foreground $r$-band luminosity density of galaxy 
clusters in SDSS DR3.  For each cluster of galaxies identified in the
SDSS DR3 redshift survey, we measure the integrated luminosity density of
galaxies out to the cluster redshift in a conical volume of radius $2^{\circ}$.
The histogram shows the number distribution of these values (in units of
$L_*$).  The blue and red arrows represent these values for the clusters
that are within $2^{\circ}$ of the UHECR cluster candidate.
} 
\label{f4}
\end{figure}

Just how unusual is this likely merging pair of galaxy clusters?  Roughly 20\% of our galaxy clusters are candidates for being in a merging pair, defined as being separated by less than 1.1 times the sum of their estimated virial radii.  With the definition of a galaxy cluster described above, there are 256 galaxy clusters containing 8.5\% of the galaxies in the sample.   The merging pair of galaxy clusters in the direction of the UHECR events has a total estimated mass of $3.1 \times 10^{14} \hMsun$; 24 \% of the 256 clusters have a mass at least this high, as do 90\% of other candidate merging clusters.  In Figure~\ref{f4} we compare the cumulative $r$-band luminosity densities in the foreground of these clusters.  The foreground of the merging clusters in the direction of the UHECRs has a lower integrated luminosity density than all but 17\% of the clusters in our sample of 256.    

\begin{figure}[h]
\epsscale{1.0}
\plotone{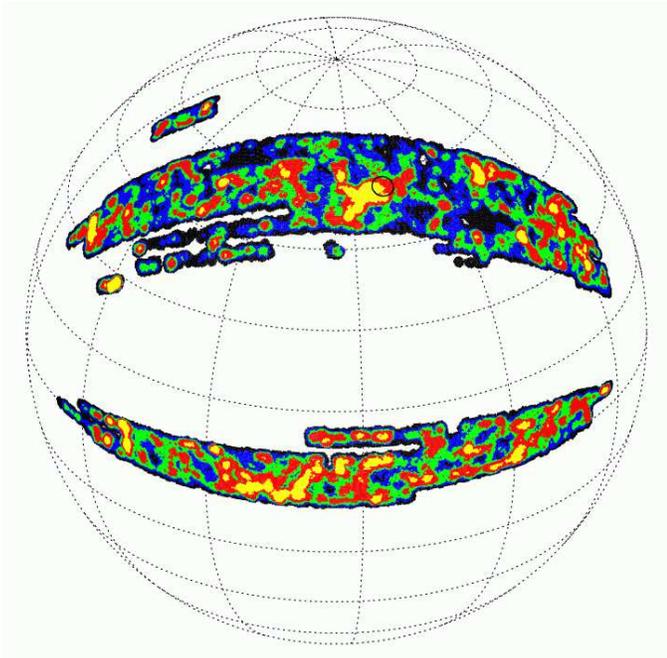}
\caption{
Integrated luminosity density map (orthographically projected) in the northern 
Galactic region of the SDSS DR3.  The luminosity density in every direction is 
averaged in a conical volume of radius $1^{\circ}$.  Colors of yellow, red, green, 
blue, and black represent luminosity densities in the upper 5, 25, 50, 75, and 95 
percentiles, respectively.  The bold black circle shows the region of radius 
$2^{\circ}$ centered on the best-fit UHECR cluster direction.
} 
\label{f5}
\end{figure}

Although merging galaxy clusters are unusual, they account for only a small fraction of the total baryonic mass and this pair of merging galaxy clusters may not be the source of the UHECR events.  In fact, as shown in Figure~\ref{f5}, the direction of the source is in the top 5 percent of all SDSS directions, from the point of view of total integrated luminosity out to 300 \hinv Mpc.  Also, the void which extends to 100 \hinv Mpc could be a red herring:  perhaps the magnetic field is generally low in all directions, and the unique aspect of this field is just its large column density of sources.  In this case, an estimate of the likely minimum distance of the source would be the distance at which the integrated luminosity density in the source direction crosses the mean for all SDSS directions, 100 \hinv Mpc.

SDSS galaxies within 4 degrees of the source include a few LINERS but no powerful AGNs.  This does not mean there are no powerful AGNs in this field since x-ray measurements have established that a large fraction of AGNs can evade detection in the optical due to obscuration.  A search of the BL-Lac catalog of \citet{veronQSO} shows that the closest BL-Lac is RXS J10586+5628, at 2.5\dg from the best-fit source direction.  The closest Egret source is also at 2.5\dg: 2EG J1054+5736.  Unless the energies of the events are mismeasured in such a way that they are much closer in energy than reported, magnetic displacement cannot shift the centroid as far as 2.5\dg.  Thus one of these objects could be the source only if there were an accidental conspiracy among angular or energy measurement errors.  

Searching the NASA Extragalactic Database within 2$^\circ$ of the quad direction, one finds few exceptional objects.  The most interesting potential source is SN1983w, in NGC3625, which is 26.2 Mpc away at RA, dec \{170.13\dg, 57.78\dg \} and thus 1.5\dg from the nominal UHECR source direction.  SN1983w was almost simultaneous with GRB831221, which appears without location in GRBCAT.  Were SN1983w a core-collapse supernova then it might be associated with the GRB and the system would be a very interesting candidate for the UHECR source.  However, this seems to be excluded on two grounds.  First, the supernova is a type 1a and therefore would not be expected to be associated with a GRB, and second, the allowed annulus of the GRB location is (just barely) incompatible with NGC3625.  We thank D. Branch and A. Fillippenko for the SN1a identification and K. Hurley for the GRB direction information.  Close associations in the BATSE catalog are irrelevant, as the photons from those GRBs arrived after the first UHECR events.  Other classes of source-types such as radio galaxies or AGNs are too abundant to find significant correlations without additional restrictions.

In summary, we have identified two unusual features in the direction of the cluster of 4-5 UHECRs:  a merging pair of galaxy clusters at $140\hmpc$ (= 200 Mpc for $h=0.7$), and a foreground void extending to 140 Mpc.  The cumulative $r$-band luminosity in the void is less than along 88\% of lines-of-sight in our SDSS sample at the same distance.  This may imply an exceptionally low magnetic field along this line of sight, which could account for the small magnetic dispersion found for these UHECRs.  \citet{gfclus} analysed the dispersion of the UHECR events to obtain $\sqrt{\langle B^2 \lambda \rangle D} \approx 1$ nG-Mpc, where $\lambda$ is the effective coherence length of the magnetic field along the line-of-sight and $D$ is the distance of the source.  The results shown in Figure~\ref{f3} strongly suggest that the source distance is at least $140$ Mpc, which implies  $\sqrt{\langle B^2 \lambda \rangle } \le 0.08~{\rm nG-Mpc}^{1/2}$ in the void out to 140 Mpc.

The merging clusters could be the locus of the source or the accelerator of the UHECRs.  The collision of the galaxy clusters may create large scale shocks capable of accelerating UHECRs\footnote{This possibility will be examined in the next round of Chandra observations, by D. Helfand and GRF.}, or the UHECRs could have been produced by an object in the merging clusters (or elsewhere along the line of sight).  No striking individual source such as a BLLac or ultra-powerful radio galaxy is evident within about 2.5\dg of the estimated UHECR source direction, although a large fraction of such sources are known to be obscured.  The region has received hardly any study at x-ray and IR wavelengths.  The density of star formation is significantly enhanced in the merging galaxy clusters, presumably leading to a corresponding enhancement in the rate of GRBs, magnetar births, and other cataclysmic events.  We provisionally conclude that the most likely distance of the UHECR source is about 140 - 210 Mpc.  We also conclude that the almost pointlike quality of this UHECR cluster may be due to exceptionally-low magnetic dispersion thanks to a long foreground void in that direction.  An examination of 2dF for similar conditions of void followed by very high density in the Auger field of view is underway.

\acknowledgements

We have benefited from information and advice from S. Balberg, M. Blanton, D. Branch, B. Dingus, A. Fillippenko, A. Fruchter, D. Helfand, K. Hurley, E. Pierpaoli, H-W. Rix and M. Shara.  Funding for the creation and distribution of the SDSS has been provided by the  Alfred P. Sloan Foundation, the Participating Institutions, NASA, the NSF, the U.S. Department of Energy, the Japanese Monbukagakusho, and the Max Planck Society. The research of G. R. Farrar has been supported in part by NSF-PHY-0401232 and NASA NAG5-9246, that of A. Berlind by NSF-PHY-0101738 and NASA NAG5-9246, and that of D. Hogg by NASA NAG5-11669 and NSF AST-0428465.


\end{document}